\documentclass[final,3p,times]{elsarticle}
\pdfoutput=1

\usepackage{lineno,hyperref}
\modulolinenumbers[5]
\usepackage{amssymb}

\usepackage{amsmath}
\usepackage{graphicx}
\usepackage{color} 

\usepackage{subcaption}
\usepackage{multirow}
\usepackage[lined,ruled,norelsize,linesnumbered]{algorithm2e}

\newcommand{\Tt}[1]{\mathbf{#1}}

\begin{document}

\begin{frontmatter}



\title{A stencil-based implementation of Parareal in the C++ domain specific embedded language STELLA}


\author[a]{Andrea Arteaga}

\author[b]{Daniel Ruprecht\corref{cor}}
\ead{daniel.ruprecht@usi.ch}

\author[b]{Rolf Krause}

\address[a]{Center for Climate System and Modeling, ETH Z\"urich, Universit\"atstrasse 16, CH-8092 Zurich, Switzerland }
\address[b]{Institute of Computational Science, Universit{\`a} della Svizzera italiana, Via Giuseppe Buffi 13, CH-6900 Lugano, Switzerland}

\cortext[cor]{Corresponding author; Phone: +41 58 666 4975}

\begin{abstract}
In view of the rapid rise of the number of cores in modern
supercomputers, time-parallel methods that introduce concurrency along
the temporal axis are becoming increasingly popular.  For the solution
of time-dependent partial differential equations, these methods can
add another direction for concurrency on top of spatial
parallelization.  The paper presents an implementation of the
time-parallel Parareal method in a C++ domain specific language for
stencil computations (STELLA).  STELLA provides both an OpenMP and a
CUDA backend for a shared memory parallelization, using the CPU or GPU
inside a node for the spatial stencils. Here, we intertwine this
node-wise spatial parallelism with the time-parallel
Parareal. This is done by adding an MPI-based implementation of
Parareal, which allows us to parallelize in time across nodes.  The
performance of Parareal with both backends is analyzed in terms of
speedup, parallel efficiency and energy-to-solution for an
advection-diffusion problem with a time-dependent diffusion
coefficient.
\end{abstract}


\end{frontmatter}


%
%

\section{Introduction}
For the numerical solution of initial value problems on massively parallel computers, concurrency is typically introduced by distributing the spatial degrees of freedom of the solution over multiple nodes, processors or cores.
In the case of time-dependent partial differential equations, this decomposition is usually induced by a decomposition of the underlying spatial computational domain and therefore this approach is referred to as ``spatial parallelism''.
The approximate solution is then computed by advancing forward in time from some given initial value by time marching methods, e.g. Runge-Kutta or multi-step schemes, which compute time-step after time-step in a serial fashion.

As the number of cores and processors in state-of-the-art supercomputers continues to rise quickly, the time dimension is more and more becoming a serial bottleneck in at least the following sense: If, for a given problem, the resolution is increased, the increase in computational cost \emph{per time-step} can be compensated by using more cores for the  parallelization of the spatial problem ---at least, as long as the used method and its implementation show good weak scaling behavior.
When the temporal resolution is increased in order to match, e.g.,  stability or accuracy constraints, however, obviously more time-steps have to be computed to reach the same final time.
Here, the resulting increase in computational cost cannot be reduced by spatial parallelization.
As discussed in~\cite{HortonEtAl1995}, the parallel complexity of time-stepping is always at least $\Theta(N_{t})$, with $N_t$ being the number of time-steps. Following~\cite{HortonEtAl1995}, $\Theta(x)$ denotes ``a positive quantity whose leading order term is proportional to $x$''.  Because of the eventual saturation of spatial strong scaling and the resulting minimum time required per time-step, the lower constant in the estimate is bounded away from zero as the number of processors increases and the minimum computing time per step is approached.

Hence, methods that introduce some degree of concurrency along the time axis are becoming increasingly popular.
Very early ideas go back to the 1960's~\cite{Nievergelt1964,MirankerLiniger1967}.
In the 1980's and 1990's, parabolic and space-time multigrid methods~\cite{Hackbusch1984,Horton1992} and parallel multiple shooting for initial value problems~\cite{Kiehl1994} were studied.
A more recent and today very widely studied method is Parareal~\cite{LionsEtAl2001}, which is also the method this paper is concerned with.
However, a number of other methods exist, e.g. the "parallel implicit time algorithm" PITA~\cite{FarhatEtAl2003}, "revisionist deferred corrections" or RIDC~\cite{ChristliebEtAl2010,ChristliebEtAl2012}, the "parallel full approximation scheme in space and time" PFASST~\cite{Minion2010,EmmettMinion2012}, space-time parallel solvers for periodic problems~\cite{ArbenzEtAl2012,ArbenzEtAl2014} and newer approaches to space-time multigrid~\cite{FriedhoffEtAl2013,Neumueller2014}.
A recent overview can be found in~\cite{Gander2014_Review}.

There is literature available on the performance of time-parallel methods for a wide range of benchmark problems from different applications, e.g. finance~\cite{BalMaday2002}, quantum chemistry~\cite{MadayTurinici2003} or plasma-physics~\cite{SamaddarEtAl2010}.
Also, particularly for Parareal, there are a number of papers concerned with analysis of the method~\cite{Bal2005,StaffRonquist2005,GanderVandewalle2007_SISC,ReynoldsEtAl2012,GanderHairer2014} or the development of improvements or modifications~\cite{FarhatCortial2006,GanderPetcu2008,RuprechtKrause2012,DaiEtAl2013,GuettelGander2013,HautWingate2013}.

Not much literature exists studying, let alone comparing, different implementation strategies for time-parallel methods: There are obviously many different possibilities to map space-time parallelism to a computer consisting of a large number of multi-core nodes, probably also equipped with accelerators.
Small-scale benchmarks for an all-MPI implementation of Parareal with spatial parallelization are discussed in~\cite{Trindade2006}, but with somewhat inconclusive performance results.
Good performance on close to half a million cores of an implementation of PFASST and parallel multigrid in space using only MPI has been demonstrated in~\cite{RuprechtEtAl2013_SC}.
An MPI-based implementation of PFASST combined with a hybrid tree-code on more than two hundred thousand cores is studied in~\cite{SpeckEtAl2012}.
Smaller scale benchmarks for a hybrid space-time parallel approach, combining OpenMP in time and MPI in space, are reported in~\cite{HaynesOng2014} for RIDC and in~\cite{RuprechtKrause2014_DDM} for Parareal.
In~\cite{Barker2013}, the original method by Nievergelt~\cite{Nievergelt1964} is implemented on GPUs.
As performance metric, all publications so far consider only speedup, other metrics like energy-to-solution, memory footprint etc. have not yet been analyzed for time parallelization.
Some concerns about the memory requirements of time-parallel methods on possible Exascale computers have been voiced in~\cite{Keyes2011}.

In this paper, we introduce a stencil-based implementation of Parareal in the recently developed C++ domain specific embedded language STELLA (STEncil Loop LAnguage)~\cite{stella}.
STELLA provides functionality for the application of finite difference stencil operators on structured grids, including an OpenMP and CUDA backend to run the computations on a multi-core CPU or a GPU.
Stencil computation on structured grids is one of the "seven dwarfs" of high-performance computing defined by Collela in 2004, see e.g.~\cite{AsanovicEtAl2009}.
The backends parallelize the stencil computation in a node across either its cores or by using an attached GPU.
The implementation of Parareal in STELLA is then employed to parallelize in time across nodes: Each node holds one "timeslice" in Parareal and performs the corresponding computations exploiting the spatial concurrency from the corresponding STELLA backend while MPI is used to communicate between nodes.
In order to optimize data transfer between GPUs, the CUDA-aware MPI implementation MPICH by Cray was used, which supports \emph{GPUDirect}. This technology avoids the need for copying the data from the GPU memory into a temporary buffer of the host memory and transfers instead the data directly to the network interface, thus reducing the latency and increasing the bandwidth.
In summary, we employ a paradigm combining a MPI-based distributed memory parallelization in time with an OpenMP or CUDA based shared memory parallelization in space.
Performance of Parareal based on both the CPU and the GPU backend is investigated in terms of speedup, parallel efficiency and energy-to-solution and compared to theoretically predicted values.
To our knowledge, this is the first time that the performance of a parallel-in-time method in terms of energy-to-solution is analyzed.

\section{Methods and implementation}\label{sec:methods}
Below, we comment briefly on serial time-stepping and then describe Parareal, including a theoretical model for expected speedup and energy overhead.
A short introduction of the STELLA language and the corresponding stencil-based implementation of Parareal is given.

\subsection{Time-marching}
Consider an initial value problem of the form
\begin{equation}
	\label{eq:ivp}
	u_{t} = f(u(t), t), \quad u(0) = u_0 \in \mathbb{R}^{d}, \quad 0 \leq t \leq T,
\end{equation}
where, in the examples below, the right hand side $f$ stems from the spatial discretization of a partial differential equation.
To fix notation for the introduction of Parareal, let the time-interval $[0,T]$ be decomposed into $N_{\rm p}$ so-called "time-slices" $[t_n, t_{n+1}]$, $n=0, \ldots, N_{\rm p}-1$ and let $u_n$ be an approximation of the solution $u$ of~\eqref{eq:ivp} at $t_n$, that is $u_n \approx u(t_n)$. Note that for Parareal, $N_{\rm p}$ is identical to the number of processors used \emph{in time}. It should not be confused with the number of processors \emph{in space} or the total number of processors.

We denote by $\mathcal{F}_{\delta t}$ a so-called ``fine integrator'', typically a higher order method with a fine time-step size $\delta t$.
That is, for some given approximation $u_n$ at $t_n$, denote by
\begin{equation}
	\label{eq:fine}
	u_{n+1} = \mathcal{F}_{\delta t}(u_n, t_{n+1}, t_{n})
\end{equation}
the approximation provided by running the fine method over the time-slice $[t_n, t_{n+1}]$ using $u_n$ as starting value.
For convenience, we assume here that $\delta t$ is such that integration over a time-slice can be done in an integer number of steps.
Otherwise, one, e.g., would have to perform a single smaller step at the end to provide $u_{n+1}$.
Evaluating~\eqref{eq:fine} step-by-step for $n=0,\ldots,N_{\rm p}-1$ corresponds to the ``classical'' application of $\mathcal{F}_{\delta t}$ as a time-marching scheme.

\subsubsection{Serial runtime}
We denote by $N_t$ the total number of fine steps, by $N_f = N_t / N_p$ the fine steps per time-slice and by $\tau_{\rm f}$ the runtime for a single step of the fine propagator.
Then, the total runtime for the serial computation of $u_{N_p}$ using $\mathcal{F}_{\delta t}$ is 
\begin{equation}
	\label{eq:cost_serial}
	C_{f} = N_{\rm t} \tau_{\rm f} =  N_{\rm p} N_{\rm f} \tau_{\rm f}.
\end{equation}
Parallelization in space, for a problem of fixed size, corresponds to reducing the computational time per time-step $\tau_{\rm f}$.
However, because strong scaling eventually saturates as communication time becomes dominant, there is a minimum value $\tau_{\rm f}^{-} \leq \tau_{\rm f}$.
The upper bound $\tau_{\rm f}^{+}$ corresponds to a serial run using only a single core.
Note that due to memory requirements such a run might often not be feasible in practice.
Hence, for any number of processors used to parallelize in space, it is
\begin{equation}
	\tau_{\rm f}^{-} \leq \tau_{\rm f} \leq \tau_{\rm f}^{+}
\end{equation}
and therefore the parallel complexity of serial time-stepping is $\Theta(N_{\rm t})$, cf.~\cite{HortonEtAl1995}.

\subsection{Parareal}
Parareal replaces serial time-marching by an iterative scheme that offers some concurrency along the time axis.
To this end, a so-called ``coarse integrator'' denoted as $\mathcal{G}_{\Delta t}$ is introduced, using a time-step $\Delta t$.
Typically, $\mathcal{G}_{\Delta t}$ will be of lower order than $\mathcal{F}_{\delta t}$ and $\Delta t \gg \delta t$.
Parareal now replaces~\eqref{eq:fine} by the iteration
\begin{equation}
	\label{eq:parareal}
	u_{n+1}^{k+1} = \mathcal{G}_{\Delta t}(u_{n}^{k+1}, t_{n+1}, t_{n}) + \mathcal{F}_{\delta t}(u_{n}^{k}, t_{n+1}, t_{n}) - \mathcal{G}(u_{n}^{k}, t_{n+1}, t_{n})
\end{equation}
with $n=0, \ldots, N_{\rm p}-1$ and $k$ the iteration counter until some convergence criterion is reached.
The initial guesses $u_n^0$ are typically generated by one start-up run of $\mathcal{G}_{\Delta t}$.
The key point is that when computing the new iterates $u_n^{k+1}$ out of given values $u^k_n$ in~\eqref{eq:parareal}, the computationally expensive evaluations of $\mathcal{F}_{\delta t}$ can be parallelized: The time required to compute all $N_{\rm p}$ values $\mathcal{F}_{\delta t}(u_n^k, t_{n+1}, t_{n})$ is then effectively equal to only a single evaluation of $\mathcal{F}_{\delta t}$.
Before computing $\mathcal{G}_{\Delta t}(u^{k+1}_{n}, t_{n+1}, t_{n})$, a process has to wait until $u^{k+1}_{n}$ is available from the processor computing the previous time-slice.
However, if the execution of the processes is pipelined properly, the effective wall clock time required for the coarse method in each iteration is only the cost of one call to $\mathcal{G}_{\Delta t}$, see~\cite{Minion2010}.

An MPI implementation of the full algorithm is sketched in Algorithm~\ref{alg:parareal}.
First, every processor computes its own initial guess $u_n^0$ by repeatedly propagating the initial value $u_0$ with the coarse propagator.
Here, processors handling later time slices have to perform more runs with the coarse method, leading to the ``pipelined'' execution discussed e.g.\ in~\cite{Minion2010} or~\cite{BerryEtAl2012}.
After the initialization phase, a fixed number of Parareal iterations is performed, but the corresponding loop could alternatively be terminated by checking some convergence criterion.
In each iteration, the processor first runs the fine propagator over its assigned time slice to produce $\tilde{u}^{k}_{p+1}$.
Then, it receives the updated initial value $u^{k+1}_{p}$ from its predecessor which, at this point, should already have completed its correction in iteration $k$ and reached the send routine.
The processor then applies the coarse propagator to the just received new initial value and computes its own updated final value $u^{k+1}_{p+1}$ according to~\eqref{eq:parareal}. 
This is then sent to the next processor, allowing it to perform its own correction and so on.
Note that process $p=0$ does not call the MPI library to receive but uses the initial value $u_0$ as starting value $u^{k+1}_{p}$ in all iterations.
Accordingly, process $p=N_{\rm p}-1$ does not send but e.g.\ writes the final value to disk.
\begin{algorithm}[th]
  \SetKwComment{Comment}{\# }{}
  \SetCommentSty{textit}
  \DontPrintSemicolon

  \KwData{Initial condition $u_0$}
  \KwResult{Output}

  \Comment{Initialization for Parareal: Perform $p-1$ runs of the coarse method to produce an initial guess $u^0_p$} 
    $u^{0}_{p} = u_0$

\For{$n = 0 \ldots p-1$}{
	$u^{0}_{p} = \mathcal{G}_{\Delta t}(u^{0}_{p}, t_{n+1}, t_{n})$ \label{alg_para_coarse1}
}

$\tilde{u}^{0}_{p+1} = \mathcal{G}_{\Delta t}(u^{0}_{p}, t_{p+1}, t_{p})$ \label{alg_para_coarse2}

  \BlankLine
  \Comment{Parareal iteration on process $p$.}
   
  \For{$k = 0 \ldots k_{max}-1$}{
	\Comment{Fine propagator}
	$\hat{u}^{k+1}_{p+1} = \mathcal{F}_{\delta t}(u^k_p, t_{p+1}, t_{p})$ \label{alg_para_fine}
	
    \BlankLine
	\Comment{Receive $u^{k+1}_{p}$ from predecessor; for $p=0$, set $u^{k+1}_{0} = u_0$.}
	\eIf {$p=0$} {
	$u^{k+1}_0 = u_0$ }
	{ MPI\_Recv$\left(u^{k+1}_{p}, p-1\right)$\label{alg_para_recv}}
	
    \BlankLine
    	\Comment{Coarse propagator}
	$\tilde{u}^{k+1}_{p+1} = \mathcal{G}_{\Delta t}(u^{k+1}_{p}, t_{p+1}, t_{p})$ \label{alg_para_coarse3}

  \BlankLine
	\Comment{Correction.}	
       $u_{p+1}^{k+1} = \tilde{u}_{p+1}^{k+1} + \hat{u}_{p+1}^{k+1} - \tilde{u}_{p+1}^{k}$ \label{alg_para_corr}
	
    \BlankLine
	\Comment{Send updated value forward; last process does not send.}
	\If {$p<N_{\rm p}-1$}{
	MPI\_Send$\left(u^{k+1}_{p+1}, p+1\right)$\label{alg_para_send}
	}
  }

  \BlankLine
  \caption{Parareal on a time slice $[t_p, t_{p+1}]$ handled by an MPI process with rank $p$.}
  \label{alg:parareal}
\end{algorithm}

\subsubsection{Parallel runtime}\label{subsubsec:parallel_runtime}
Assume that $\Delta t$ is chosen such that $\mathcal{G}_{\Delta t}$ integrates over one time-slice in an integer number of steps $N_{\rm c}$ and denote by $\tau_{\rm c}$ the cost of a single coarse time-step.
The initialization phase then needs $N_{\rm p} N_{\rm c}$ many time steps on the last processor, where it takes the longest.
After that, each Parareal iteration costs $N_c$ coarse steps and $N_f$ fine steps, where, for the sake of simplicity, we ignore overhead from e.g.\ communication.
In total, the expected time-to-solution for the pipelined version of Parareal with $K$ iterations and one serial run of $\mathcal{G}_{\Delta t}$ for start-up is then
\begin{equation}
	\label{eq:cost_parareal}
	C_{\rm p} =  N_{\rm p} N_{\rm c} \tau_{\rm c} + K \left( N_{\rm c} \tau_{c} +  N_{\rm f} \tau_{f} \right) = \left( N_{\rm p} + K \right) N_{\rm c} \tau_c + K N_{\rm f} \tau_{f}.
\end{equation}
Note that if the total number of fine time steps $N_{\rm t} = N_{\rm p} N_{\rm f}$ is increased together with the number of processors $N_{\rm p}$, so that the number of fine steps per slice $N_{\rm f}$ remains constant, the effective cost for the serial method in Parareal, i.e.~the second term in~\eqref{eq:cost_parareal}, stays the same -- at least as long as the number of iterations $K$ does not increase.
This, however, is typically not the case, but as shown in~\cite{GanderVandewalle2007_SISC}, for diffusive problems the increase in iterations with increasing $N_{\rm p}$ is small.
Therefore, if a cheap enough coarse method can be found that still leads to good convergence, the cost from having to compute more time steps increases only slowly through the first term in~\eqref{eq:cost_parareal} and a mild increase of $K$.
Note that recent modifications of Parareal towards a full-fledged multigrid in time show no increase in iteration count, at least for simple diffusive problems~\cite{FriedhoffEtAl2013}.

\subsubsection{Speedup}\label{subsubsec:speedup}
For a fixed resolution, Parareal provides an additional direction of concurrency and can extend strong scaling beyond the saturation of the spatial parallelization:
Combining~\eqref{eq:cost_parareal} and~\eqref{eq:cost_serial} provides a theoretical bound for the speedup for Parareal 
\begin{equation}
	\label{eq:speedup}
	S_{\rm bound}(N_{\rm p}) = \frac{C_{\rm f}}{C_{\rm p}} =  \frac{1}{ \left( 1 + \frac{K}{N_{\rm p}} \right) \frac{N_{\rm c}}{N_{\rm f}} \frac{\tau_{\rm c}}{\tau_{\rm f}} + \frac{K}{N_{\rm p}}},
\end{equation}
see again also~\cite{Minion2010}.
As the number of processors $N_{\rm p}$ increases, while $N_{\rm f}$ and $N_{\rm c}$ remain fixed, two different bounds on the maximum speedup obtainable by Parareal can be derived from~\eqref{eq:speedup}: 
\begin{equation}
	S_{\rm bound}(N_{\rm p}) \leq \frac{N_{\rm p}}{K} \quad \text{and} \quad S_{\rm bound}(N_{\rm p}) \leq \frac{N_{\rm f}}{N_{\rm c}} \frac{\tau_{\rm f}}{\tau_{\rm c}}.
\end{equation}
Neglecting the first term in the denominator in~\eqref{eq:speedup} yields the first, neglecting the second term yields the second bound.
The first bound illustrates that good performance of Parareal requires rapid convergence in only a few iterations: The maximum parallel efficiency is bounded by $1/K$.
The second bound stems from the ratio of the runtime of the fine to the coarse method: The latter has to be significantly cheaper in order to allow for reasonable speedup.
Using time steps $\Delta t \gg \delta t$ improves the ratio $N_{\rm f}/N_{\rm c}$.
The ratio $\tau_{\rm f}/\tau_{\rm c}$ can be improved e.g.\ by using a lower order method for $\mathcal{G}_{\Delta t}$, a lower order spatial discretization~\cite{RuprechtKrause2012} or even a coarser spatial mesh~\cite{FischerEtAl2005}. 
The latter approach, however, requires interpolation and restriction operators between the two spatial meshes and the order of interpolation can have a significant influence on convergence~\cite{Ruprecht2014_GAMM}.

\paragraph{Parallel Efficiency}
Ideally, a parallel code using $N_{\rm p}$ processors would run $N_{\rm p}$ times faster than its serial counterpart, therefore providing ideal speedup
\begin{equation}
	S_{\rm ideal}(N_{\rm p}) = N_{\rm p}.
\end{equation}
For Parareal, however, ideal speedup is by design not possible and the achievable speedup is limited by $S_{\rm bound}(N_{p})$.
Given some measured speedup $S_{\rm measured}(N_{\rm p})$, we can therefore consider and compare the following two parallel efficiencies
\begin{equation}
	\label{eq:efficiency}
	E_{\rm bound} \left(N_{\rm p} \right) = \frac{S_{\rm bound} \left(N_{\rm p}\right)}{S_{\rm ideal}(N_{\rm p})} \quad \mbox{and} \quad E_{\rm measured}\left(N_{\rm p}\right) = \frac{S_{\rm measured}\left(N_{\rm p}\right)}{S_{\rm ideal} \left(N_{\rm p}\right)}.
\end{equation}
Here, $E_{\rm bound}$ is the maximum theoretically possible efficiency for Parareal while $E_{\rm measured}$ is the measured efficiency.
Ideally, we want $E_{\rm measured}$ to closely match $E_{\rm bound}$.

\paragraph{Energy consumption}
Denote by $Q_{\rm s}$ the energy in Joule consumed by a serial run on a single node with runtime $T_{\rm s}$ (in seconds) and by $Q_{\rm p}$ a parallel run on $N_{\rm p}$ nodes with runtime $T_{\rm p}$.
The corresponding powers in Watt, equal to Joule per second, that is energy per unit time, can approximately be computed from
\begin{equation}
	P_{\rm p} = \frac{Q_{\rm p}}{T_{\rm p}}, \quad \mbox{and} \quad P_{\rm s} = \frac{Q_{\rm s}}{T_{\rm s}}.
\end{equation}
In an ideal case, operating $N_{\rm p}$ nodes would require exactly $N_{\rm p}$ times as much power as operating a single node, so that
\begin{equation}
	P_{\rm p} = N_{\rm p} P_{\rm s}.
\end{equation}
Therefore, the ideal ratio of overhead in terms of consumed energy would be
\begin{equation}
	\gamma_{\rm ideal} = \frac{Q_{\rm p}}{Q_{\rm s}} = \frac{N_{\rm p}}{S_{\rm p}} = E_{\rm p}^{-1}
\end{equation}
where
\begin{equation}
	S_{\rm p} = \frac{T_{\rm s}}{T_{\rm p}}, \quad \text{and} \quad E_{\rm p} = \frac{S_{\rm p}}{N_{\rm p}}
\end{equation}
denote speedup provided by $N_{\rm p}$ nodes for the parallel run and the resulting parallel efficiency.
We thus expect the energy ratio of the parallel to the serial run to be approximately the inverse of the parallel efficiency.
Note that for a method with $100\%$ efficiency, that is $E_{\rm p} = 1$, there will be no overhead so that $\gamma_{\rm ideal} = 1$.
Parareal, however, cannot provide optimal efficiency by design.
In order to distinguish between energy overhead as a result of intrinsic suboptimal scaling and additional overhead e.g. from communication, define the intrinsic minimal energy overhead for Parareal as
\begin{equation}
	\label{eq:gamma_expected}
	\gamma_{\rm bound} = \frac{N_{\rm p}}{S_{\rm bound}}
\end{equation}
according to the maximum theoretically possible speedup $S_{\rm bound}$.
In Section~\ref{sec:results} we will demonstrate that measured energy overhead of Parareal matches $\gamma_{\rm bound}$ reasonably well.

\subsubsection{Convergence}\label{subsubsec:conv_parareal}
To assess convergence of Parareal, we use the relative defect between the solution at the final time $T = t_{N_{\rm p}}$ provided by Parareal after $k$ iterations $u^{k}_{N_{\rm p}}$ and the solution $u_{\rm fine}$ provided by running the fine method serially
\begin{equation}
	\label{eq:defect}
	d^{k} := \frac{ \left\| u^{k}_{N_{\rm p}} - u_{\rm fine} \right\|_{\infty} }{ \left\| u_{\rm fine} \right\|_{\infty} }.	
\end{equation}
Note that $u_{\rm fine}$ is obtained by evaluating~\eqref{eq:fine} step-by-step for $n=0, \ldots, N_{\rm p}-1$.
The relative discretization error of the solution provided by Parareal can be estimated as
\begin{equation}
	\varepsilon_{\rm parareal} := \frac{ \left\| u^{k}_{N_{\rm p}} - u_{\rm exact} \right\|_{\infty} }{ \left\| u_{\rm exact} \right\|_{\infty} } \leq \frac{ \left\| u^k_{N_{\rm p}} - u_{\rm fine} \right\|_{\infty} }{ \left\| u_{\rm exact} \right\|_{\infty} } + \frac{ \left\| u_{\rm fine} - u_{\rm exact} \right\|_{\infty} }{ \left\| u_{\rm exact} \right\|_{\infty} } = d^{k} \frac{\left\| u_{\rm fine} \right\|_{\infty} }{ \left\| u_{\rm exact} \right\|_{\infty} } + \varepsilon_{\rm fine}
\end{equation}
where $\varepsilon_{\rm fine}$ is the relative discretization error of the serial fine method.
Because $\left\| u_{\rm fine} \right\| / \left\| u_{\rm exact} \right\| \approx 1$, if $d^k \ll \varepsilon_{\rm fine}$, we get $\varepsilon_{\rm parareal} \approx \varepsilon_{\rm fine}$, that is the Parareal solution has essentially the same discretization error as the serial fine integrator.

Note that~\eqref{eq:defect} can of course only be computed in benchmark simulations, where the fine solution is computed, too.
In production runs, where one replaces the fine propagator by Parareal, other means have to be used to monitor Parareal's convergence: Possible choices would be the difference between two consecutive iterates, i.e. $\left\| u^{k+1}_{p} - u^{k}_{p} \right\|$ or the residual $\left\| \mathcal{F}_{\delta t}(y^{k}_{p-1}) - y^{k}_{p} \right\|$ at each time slice, cf.~\cite{Ruprecht2014_GAMM}.

\subsection{Stencil language}\label{subsec:stencil}
\begin{figure}[t]
	\centering
	\includegraphics[width=0.50\textwidth]{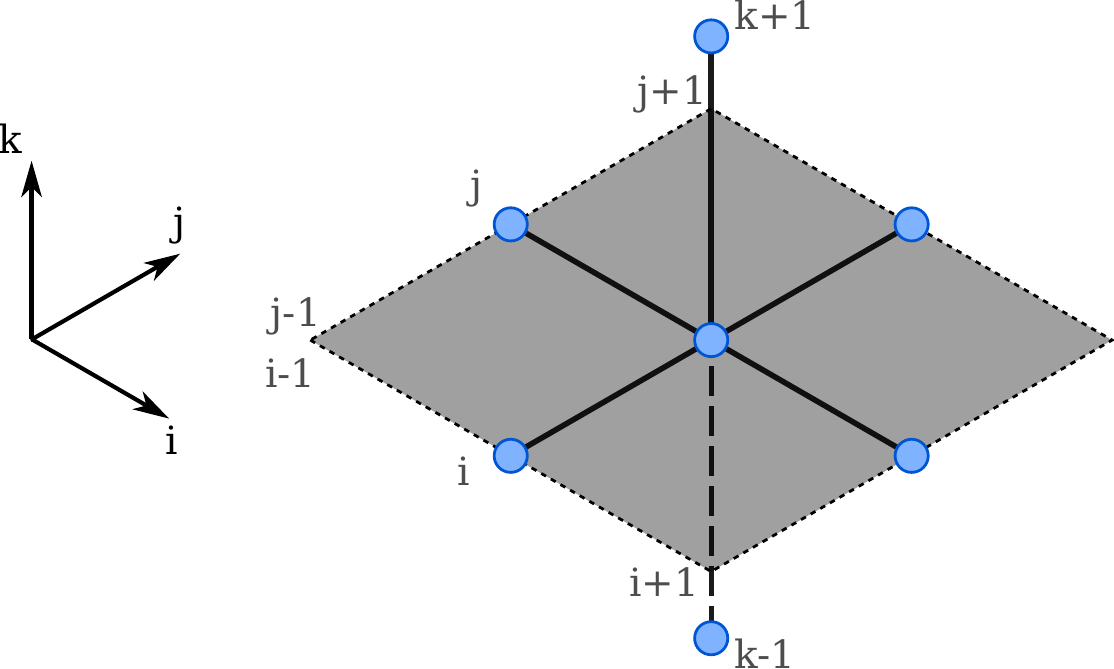}
	\caption{Diagram of the access pattern of a stencil. In this example the neighboring points around the central point $u_{i,j,k}$ are accessed for the computation. This pattern is used for instance in the computation of the Laplacian of a three-dimensional field. \label{fig:stencil}}
\end{figure}
STELLA (STEncil Loop Language) is a C++ domain specific embedded language designed and written by T. Gysi et al. \cite{stella} within the Swiss HP2C programme (High Performance and High Productivity Computing).
It allows the user to define abstract formulations for stencil formulas and deploys the computation in a high-performance environment by using multiple threads or graphical processors (GPUs).
The framework optimizes the data layout, the memory accesses, the concurrency schemes and many other factors for optimal performance.
An extensive description can be found in STELLA's documentation \cite{gysistella}.
Although the main motivation behind the development of STELLA is the easy and efficient implementation of finite-difference stencils, its application domain is more general.
A stencil computation is defined as any operation applied identically on all grid cells of a multi-dimensional array that requires a local set of neighboring values. 
The access pattern can be represented in a diagram where every point corresponds to a value used in the computation. 
Figure \ref{fig:stencil}, for instance, shows the access pattern of a computation that requires all the nearest neighboring values of the point where the operation is applied. 
One example of such an operator is the second-order discretization of the Laplacian of a three-dimensional field
\begin{equation}
 \Delta u_{i,j,k} \approx \frac{u_{i+1,j,k} + u_{i-1,j,k} + u_{i,j+1,k} + u_{i,j-1,k} + u_{i,j1,k+1} + u_{i,j,k-1} - 6 u_{i,j,k}}{\Delta x ^ 2}.
\end{equation}
Again for the sake of simplicity, we assume here that step size in all dimensions are identical, i.e. $\Delta x = \Delta y = \Delta z$, but this is not an intrinsic limitation of STELLA.
However, not just finite-difference operators can be considered stencils, but also other computations fall into this category. For instance, the simple sum of two fields with a multiplicative factor can be considered a stencil and thus executed by STELLA:
\begin{equation}
u_{i,j,k} \; +\!\!= \; \alpha \cdot v_{i,j,k},
\end{equation}
where $\left( a +\!\!=  b \right) := \left( a = a+b \right)$ following C++ syntax.
The access pattern diagram of such an operation would be a simple point at the position $(i,j,k)$.

Here, we introduce a stencil-based implementation of Parareal, using STELLA's backends to parallelize loops over the vector of degrees-of-freedoms and benchmark it for the three-dimensional advection diffusion equation~\eqref{eq:adv_diff_eq} introduced below in Section~\ref{sec:results}.
In particular, STELLA parallelizes the evaluation of the finite difference stencils in the spatial discretization, providing us with easy access to shared-memory spatial parallelism.
We employ STELLA for the following tasks in Parareal, sketched in~Algorithm~\ref{alg:parareal}:
\begin{itemize}
\item Coarse propagation (Algorithm~\ref{alg:parareal}, Lines~\ref{alg_para_coarse1}, \ref{alg_para_coarse2} and \ref{alg_para_coarse3}). 
Here, we use for $\mathcal{G}_{\Delta t}$ a forward Euler method with a first order upwind discretization of the convective term $\Tt{c} \cdot \nabla u(\Tt{x},t)$ in~\eqref{eq:adv_diff_eq} and a second order centered discretization of the diffusive term $\nu(t) \Delta u(\Tt{x},t)$.
Algorithm~\ref{alg:coarse} sketches one call of $\mathcal{G}_{\Delta t}$, performing $N_c$ steps with the coarse method.
STELLA performs two different stencil computations in each step:
The computation of the right hand side (Algorithm~\ref{alg:coarse}, Lines~\ref{alg_coarse_rhsstart}-\ref{alg_coarse_rhsend}), that is the evaluation of the two finite difference stencils, is combined into a single complex STELLA stencil (for readability, both stencils are shown independently in Algorithm~\ref{alg:coarse}, though).
The for-loop at Line~\ref{alg_coarse_for} is performed and parallelized by STELLA. 
After the computation of the right hand side, the forward Euler step in Line~\ref{alg_coarse_euler} is again performed and parallelized by STELLA.
\item Fine propagation (Algorithm~\ref{alg:parareal}, Line~\ref{alg_para_fine}). Analogously to the coarse propagation, the fine propagation scheme consists of a sequence of time steps. 
It uses centered fourth order differences for both advection and diffusion and a fourth order accurate explicit Runge-Kutta method (the classical Runge-Kutta-4 scheme).
Both the spatial finite difference stencils as well as the summation of the stages in the Runge-Kutta method are done with STELLA, just as for the coarse method.
\item Correction (Algorithm~\ref{alg:parareal}, Line~\ref{alg_para_corr}). At the end of the parareal iteration, after having computed the fine solution, having received the updated solution from the previous process, and having computed the coarse solution, the algorithm must apply the update formula \eqref{eq:parareal}, which is a sum of three fields. This sum is also computed and parallelized by STELLA.
\end{itemize}
The spatial stencils sketched in Algorithm~\ref{alg:coarse} correspond to the discretization of a linear advection-diffusion equations as used in Section~\ref{sec:results} for benchmarking.
However, defining different stencils would easily allow to treat other PDEs as well.
Modifying the stencils for the stages would also allow to use other time stepping schemes.
\begin{algorithm}[h!]
\SetKwComment{Comment}{\#}{}
\SetCommentSty{textit}
\DontPrintSemicolon
\KwData{Initial value $u$}
\KwResult{$\tilde{u} = \mathcal{G}_{\Delta t}(u)$}

\BlankLine

\For{$n=0 \ldots N_{c}$}{
    \For{$i,j,k \in \mathbb{F}$}{ \label{alg_coarse_for}
    	$rhs_{i,j,k} = \nu(t) \cdot \left( u_{i+1,j,k} + u_{i-1,j,k} +u_{i,j+1,k} + u_{i,j-1,k} +u_{i,j,k+1} + u_{i,j,k-1} - 6 \cdot u_{i,j,k} \right) / (\Delta x\cdot \Delta x) $ \label{alg_coarse_rhsstart}
    	
        \eIf{$c_x > 0$}{
    	    $rhs_{i,j,k} \;\, -\!\!= \; c_x \cdot (u_{i,j,k}-u_{i-1,j,k}) / \Delta x$
    	}{
    	    $rhs_{i,j,k} \;\, -\!\!= \; c_x \cdot (u_{i+1,j,k}-u_{i,j,k}) / \Delta x$
    	}
        \eIf{$c_y > 0$}{
    	    $rhs_{i,j,k} \;\, -\!\!= \; c_y \cdot (u_{i,j,k}-u_{i,j-1,k}) / \Delta x$
    	}{
    	    $rhs_{i,j,k} \;\, -\!\!= \; c_y \cdot (u_{i,j+1,k}-u_{i,j,k}) / \Delta x$
    	}
        \eIf{$c_z > 0$}{
    	    $rhs_{i,j,k} \;\, -\!\!= \; c_z \cdot (u_{i,j,k}-u_{i,j,k-1}) / \Delta x$
    	}{
    	    $rhs_{i,j,k} \;\, -\!\!= \; c_z \cdot (u_{i,j,k+1}-u_{i,j,k}) / \Delta x$
    	}\label{alg_coarse_rhsend}
    }
    \For{$i,j,k \in \mathbb{F}$}{
    	    $\tilde{u}_{i,j,k} \; \!\!= \; u_{i,j,k} + \Delta t \cdot rhs_{i,j,k}$ \label{alg_coarse_euler}
    }
}
Return $\tilde{u}$
\caption{Explicit Euler as coarse propagator for the convection-diffusion problem~\eqref{eq:adv_diff_eq} implemented in STELLA. A first order upwind stencil is used for the convective term, a second order centered stencil for the diffusive term. The fine method consists of a similar implementation of a fourth order Runge-Kutta method with fourth-order space discretization (not shown here).}
\label{alg:coarse}
\end{algorithm}

\subsection{Mapping to hardware}
\begin{figure}[t]
	\centering
	\includegraphics[width=0.75\textwidth]{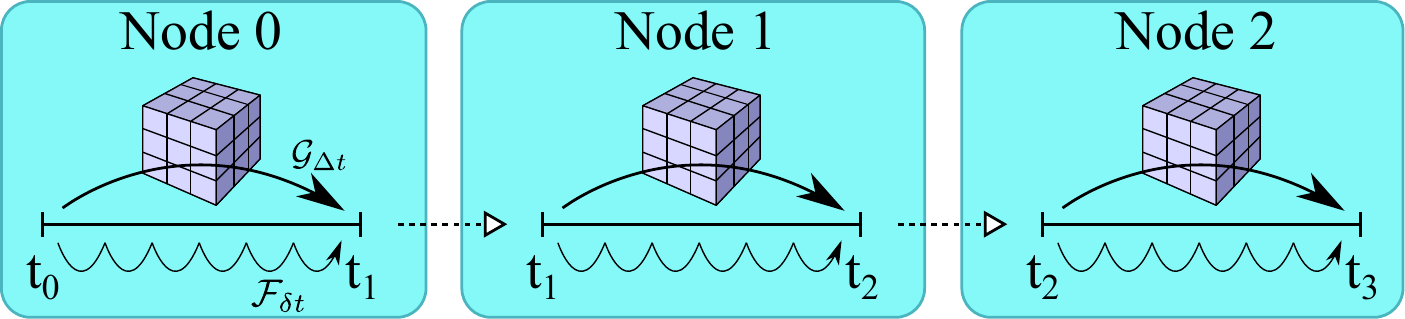}
	\caption{Mapping of Parareal-STELLA to multiple compute nodes on Piz Daint: Each of the $N_{\rm p}$ time-slices is assigned to a node. The stencil computations required to advance the solution over this time-slices are carried out by STELLA, represented by the blue cube, using the coarse integrator $\mathcal{G}_{\Delta t}$ or using the fine integrator $\mathcal{F}_{\delta t}$ (solid arrows). The dashed arrows represent the communications from one node to the other, which occur once per Parareal iteration.}\label{fig:mapping}
\end{figure}

There are many possible ways to map Parareal (or any other time-parallel method for that matter) to a specific parallel computer.
Here, we assign each time-slice in Parareal to one compute node and let STELLA make use of either the node's CPUs through its OpenMP backend or the node's GPU through its CUDA backend to parallelize the stencil computations required to run $\mathcal{G}_{\Delta t}$ and $\mathcal{F}_{\delta t}$.
Communication in time in Parareal (lines~\ref{alg_para_recv} and~\ref{alg_para_send} in Algorithm~\ref{alg:parareal}) between different nodes is done using the MPI library.
The setup is sketched in Figure~\ref{fig:mapping}: The time slice $[t_p, t_{p+1}]$ is handled by MPI process $p$, which is assigned one full compute node.
The resources of the node (multi-core CPU or GPU) are then used for the iteration sketched in Algorithm~\ref{alg:parareal}, cf. the discussion in~\ref{subsec:stencil}.
After a completed Parareal iteration, the update is sent to the next node/process using MPI.
For the CUDA backend, we use the \emph{GPUDirect mode} MPI extension, which allows for a direct communication between GPUs on different nodes without the need for temporary buffers in the system memory.
For the OpenMP backend, non-blocking communication (i.e. MPI\_IRECV, MPI\_ISEND and corresponding waits) could be used and has been tested, in order to overlap communication and computation. 
However, the benefits turned out to be negligible. 
Because non-blocking communication was unavailable for direct GPU communication at the time of writing the software, we decided to make the comparison between CPU and GPU fairer by using a blocking communication scheme for Parareal for both backends.

In summary, we here combine a distributed memory parallelization in time with shared memory parallelization in space using a multi-core CPU or a GPU.
Other approaches have been studied as well, e.g.\ a fully MPI-based space-time parallelization~\cite{RuprechtEtAl2013_SC}, a MPI-in-time parallelization with a hybrid spatial parallelization~\cite{SpeckEtAl2012}, a hybrid space-time approach combining OpenMP in time with MPI in space~\cite{RuprechtKrause2014_DDM,HaynesOng2014} or a GPU-based implementation~\cite{Barker2013}
Note that while feasibility and efficiency has been demonstrated for all these different strategies, so far there are no comparisons of their relative strengths and weaknesses.
Some implementation strategies like PGAS languages or one-sided communication have, to our knowledge, not yet been explored at all for parallel-in-time methods.

\section{Numerical results}\label{sec:results}
For the benchmarks reported below, we use the implementation of Parareal in STELLA described in Section~\ref{sec:methods} to solve the three-dimensional advection-diffusion equation with a time-dependent diffusion coefficient 
\begin{equation}
	\label{eq:adv_diff_eq}
	u_t(\Tt{x},t) + \Tt{c} \cdot \nabla u(\Tt{x},t) = \nu(t) \Delta u(\Tt{x},t), \quad 0 \leq t \leq T,
\end{equation}
with $\Tt{c} \in \mathbb{R}^{3}$ and $\nu(t) \in \mathbb{R}^{+}$.
The spatial domain is $\Omega = [0,1]^{3}$ with periodic boundary conditions. The initial value is
\begin{equation}
	u(\Tt{x},0) = u_0(\Tt{x}) = \sin(2\pi x) \sin(2\pi y) \sin(2\pi z).
\end{equation}
For now, let $\Tt{c} = 0$ and assume a solution of the form
\begin{equation}
	\label{eq:sol_form}
	u(\Tt{x},t) = a(t) u_0(\Tt{x}).
\end{equation}
Because $\Delta u_0(\Tt{x}) = -12 \pi^{2} u_0(\Tt{x})$, this results in the initial value problem
\begin{equation}
	a'(t) = -12 \pi^{2} \nu(t) a(t), \quad a(0) = 1
\end{equation}
 for the coefficient $a$ in~\eqref{eq:sol_form}.
Variation of parameters gives the solution
\begin{equation}
	a(t) = \exp\left(- \int_0^{t} 12 \pi^{2} \nu(s) \ ds \right).
\end{equation}
Note that for $\nu \equiv 1$, this reduces to $a(t) = \exp(-12 \pi^2 t)$. For a simple profile of the form
\begin{equation}
	\nu(t) = \nu_0 + \frac{\nu_0}{2} \sin(\omega t), \quad \nu_0 > 0
\end{equation}
we get
\begin{equation}
	a(t) = -12 \pi^{2} \left( \nu_0 t - \frac{\nu_0}{2 \omega} \left( \cos(\omega t) - 1 \right) \right).
\end{equation}
For an arbitrary $\Tt{c} \in \mathbb{R}$, the full solution then is obtained as
\begin{equation}
	u(\Tt{x},t) = a(t) u_0(\Tt{x} - \Tt{c} t).
\end{equation}
In the tests reported below, we use $\nu_0 = 0.1$, with $\omega = 0$ (no time-dependance in $\nu$) and $\omega = 100$ (rapidly oscillating $\nu$). 
Furthermore, the advection velocity is set to $\Tt{c} = (1,1,1)$ and the simulation is run until $T = 0.1$.

Here, to allow for precise benchmarking of STELLA-Parareal, we choose a rather simple linear problem where an analytic solution is available.
This allows to compute and compare the defect of Parareal with the discretization error of the coarse and fine propagator, cf. the discussion in~\ref{subsec:conv_parareal}.
In production runs, this solution won't be known and other means have to be used to monitor convergence of Parareal and to decide when to stop iterating, see also~\ref{subsubsec:conv_parareal}.

\subsection{Discretization}
The fine propagator $\mathcal{F}_{\delta t}$ uses fourth order centered finite differences for both advection and diffusion on a spatial mesh with $128^{3}$ points. 
A Runge-Kutta-4 method with a time-step $\delta t = \frac{T}{2^{15}}$ is used for integration, resulting in an overall relative error at the end of the simulation of $\varepsilon_{\rm fine} \approx 1.9 \times 10^{-7}$ for $\omega=0$ and of $4.8 \times 10^{-6}$ for $\omega=100$.
The coarse method $\mathcal{G}_{\Delta t}$ uses a first order upwind stencil for advection and a second order centered stencil for diffusion, again on a spatial mesh with $128^{3}$ points.
A first order forward Euler method with time-step $\Delta t = \frac{T}{2^{11}}$ is used, resulting in an overall relative error at $T=0.1$ of about $\varepsilon_{\rm coarse} \approx 4.5 \times 10^{-2}$ for both $\omega=0$ and $\omega=100$.
The time-steps are chosen such that the temporal discretization error is about the same as the spatial discretization error, i.e., further refinement in time without simultaneously refining the spatial mesh does no longer improve the accuracy of the solution.

Note that if the time step of $\mathcal{F}_{\delta}$ is close to the stability limit (and not, as in the case considered here, determined by accuracy considerations), the coarse propagator will very likely have to be an implicit method in order to allow for a coarse time step $\Delta t \gg \delta t$. Alternatively, it is possible to use similar time steps for both coarse and fine method and reduce the computational cost of $\mathcal{G}_{\Delta t}$ by other means, e.g. aggressive coarsening of the spatial discretization.

\subsection{Hardware}
All runs are performed on the Cray XC30 "Piz Daint" at the Swiss National Supercomputing Centre (CSCS).
Piz Daint consists of 5'272 hybrid compute nodes, each with an octo-core Intel Xeon Processor E5-2670 and an NVIDIA Tesla K20X GPU, connected through the high-bandwidth, low-latency Aries routing and communications ASIC.
The system provides a rich set of programming environments and software, including the GNU, Cray, Intel and PGI compilers and the Cray derivative of the MPICH2 implementation of the MPI library. In our experiments we used the GNU compiler in its version 4.8 and applied the standard optimization features provided by the switch \texttt{-O3}.

\subsection{Results}
In order to give a fair and meaningful assessment of the speedup provided by Parareal, it is important to make sure that it provides comparable accuracy as the time-serial fine solution.
Therefore, in \ref{subsec:conv_parareal}, some precursory tests are run to ensure that a configuration is found for Parareal for which reported speedups do actually compare solutions of comparable accuracy, that is $\varepsilon_{\rm fine} \approx \varepsilon_{\rm parareal}$, cf. the discussion in~\ref{subsubsec:conv_parareal}.
This aspect is sometimes not considered, but it should be noted that if the time-parallel solution does not provide comparable accuracy as running the fine method serially, reporting speedups  by comparing their runtimes becomes essentially meaningless.
Sufficiently many iterations of Parareal have to be performed in order to reduce the defect $d^k$ defined in~\eqref{eq:defect} below the relative discretization error $\varepsilon_{\rm fine}$ of the fine serial solution.
Here, we have an analytical solution at hand to compute $\varepsilon_{\rm fine}$.

Moreover, for a fair comparison of STELLA's CPU and GPU backends, the optimal number of threads to be used on one node by the CPU version is determined.
Performance results for the time parallelization are shown in~\ref{subsec:speedup} while in~\ref{subsec:energy} Parareal's efficiency in terms of energy is analyzed.

\subsubsection{Convergence of Parareal}\label{subsec:conv_parareal}
\begin{figure}[t]
	\centering
	\includegraphics{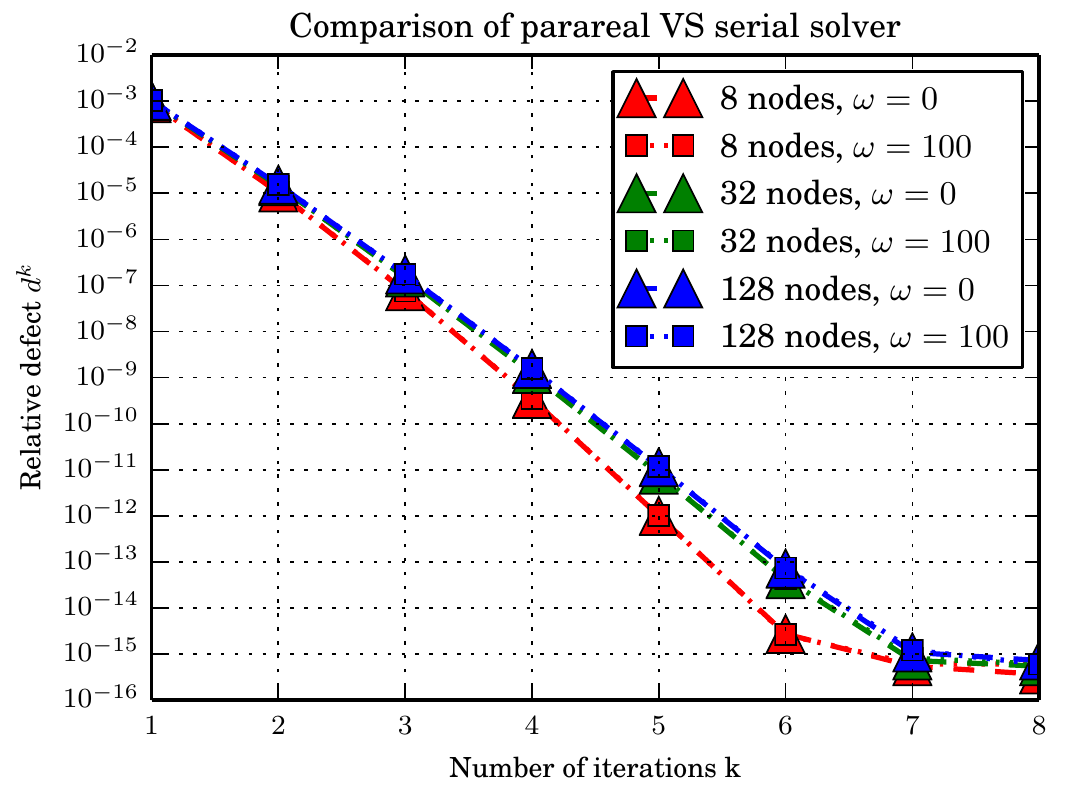}
	\caption{Relative defect $d^k$, see~\eqref{eq:defect}, between Parareal and the serial fine solution versus number of iterations. Shown are results for different numbers of nodes, that is different numbers $N_{\rm p}$ of concurrently computed time-slices, and for values of $\omega = 0$ (no time-dependance in the diffusion coefficient) and $\omega=100$ (rapidly oscillating diffusion coefficient).}\label{fig:parareal_conv}
\end{figure}
Figure~\ref{fig:parareal_conv} shows $d^k$ versus the number of iterations for $N=8$, $32$, $128$ concurrently computed time-slices and for $\omega=0$ and $\omega=100$.
Parareal converges rapidly in all cases and neither the number of time-slices nor the time-dependent diffusion coefficient has a significant impact.
This matches the analysis for constant-coefficient diffusive problems in~\cite{GanderVandewalle2007_SISC}.
The marginal affect of a time-dependent diffusion coefficient is in line with the study of Parareal's convergence for space- and time-dependent diffusion coefficients in~\cite{RuprechtEtAl2014_DDM}.
See also the comments in~\ref{subsubsec:parallel_runtime}.
Note that both the results presented here as well as the papers mentioned above are concerned with linear problems: For complex nonlinear problems, less favorable convergence behavior is possible.

After three iterations, the defect between Parareal and $\mathcal{F}_{\delta t}$ is in all cases significantly smaller than the discretization error of $\varepsilon_{\rm fine} \approx 4.8 \times 10^{-6}$ for $\omega=100$ and about the order of magnitude of the fine error for $\omega=0$.
Hence, for $k=3$ the parallel and serial solution have essentially the same error and we thus report below speedups for Parareal with $\omega=100$ and $k=3$ iterations.

\subsubsection{Optimal number of threads for STELLA on CPUs}\label{subsec:stella_threads}
\begin{figure}
	\centering
	\includegraphics{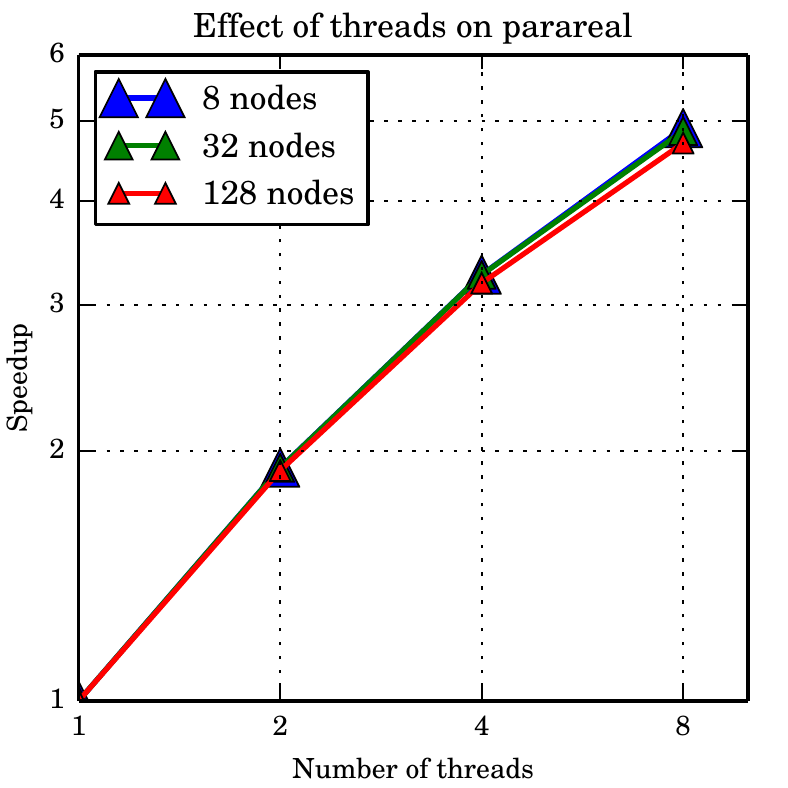}
	\caption{Speedup from the spatial parallelization using STELLA's CPU backend for multi-threaded stencil computations across different numbers of nodes used for Parareal.}\label{fig:speedup_threads}
\end{figure}
To provide a fair serial reference for the CPU version of STELLA, we have to find the optimal number of threads to be used for the stencil computation on a node.
Therefore, we run Parareal and keep the number of time-slices, and thus the number of nodes, fixed and vary the number of OpenMP threads used for the spatial parallelization in each node from 1 to 8 (because Piz Daint has 8 cores per node).

Figure~\ref{fig:speedup_threads} shows the speedup provided by using more threads for STELLA, measured against Parareal using only a single thread on each node for computation of the spatial stencils.
Although speedup from using more threads is less than optimal, it increases up to all eight cores in a node and therefore, in the CPU benchmarks reported below, STELLA's CPU backend is always using 8 threads on each node.
It is interesting to note that the speedup from STELLA seems to be hardly affected by the number of nodes used for Parareal: The speedup curves for Parareal across 8 nodes and across 128 nodes are nearly identical.
This suggests that here spatial and temporal parallelization function more or less independently and do not interfere with each other.
The reason is probably that Parareal essentially uses $\mathcal{F}_{\delta t}$ and $\mathcal{G}_{\Delta t}$ as black-boxes with very little interweaving of space- and time solver.
In case of more closely intertwined space-time parallelization, better overall speedup can be achieved, but a more complex interplay of the efficiencies of both approaches can be encountered, see~\cite{SpeckEtAl2014_Parco}.

\subsubsection{Speedup and runtimes for Parareal}\label{subsec:speedup}
\begin{figure}[t]
	\centering
	\begin{subfigure}[t]{0.495\textwidth}
		\centering
		\includegraphics[width=0.99\textwidth]{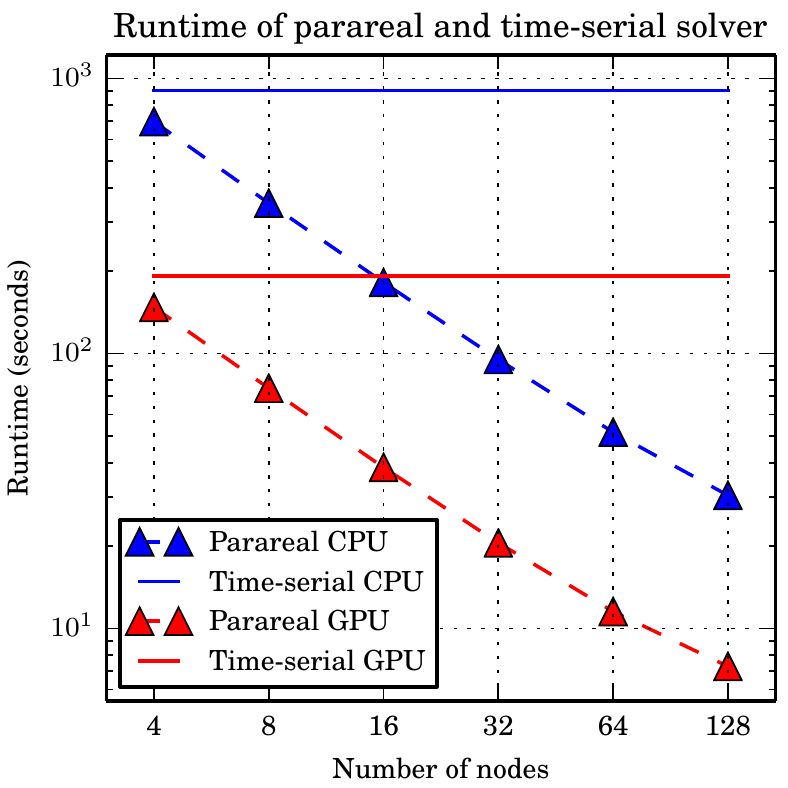}
		\caption{Runtime in seconds}\label{fig:runtimes}		
	\end{subfigure}	
	\begin{subfigure}[t]{0.495\textwidth}
		\centering
		\includegraphics[width=0.99\textwidth]{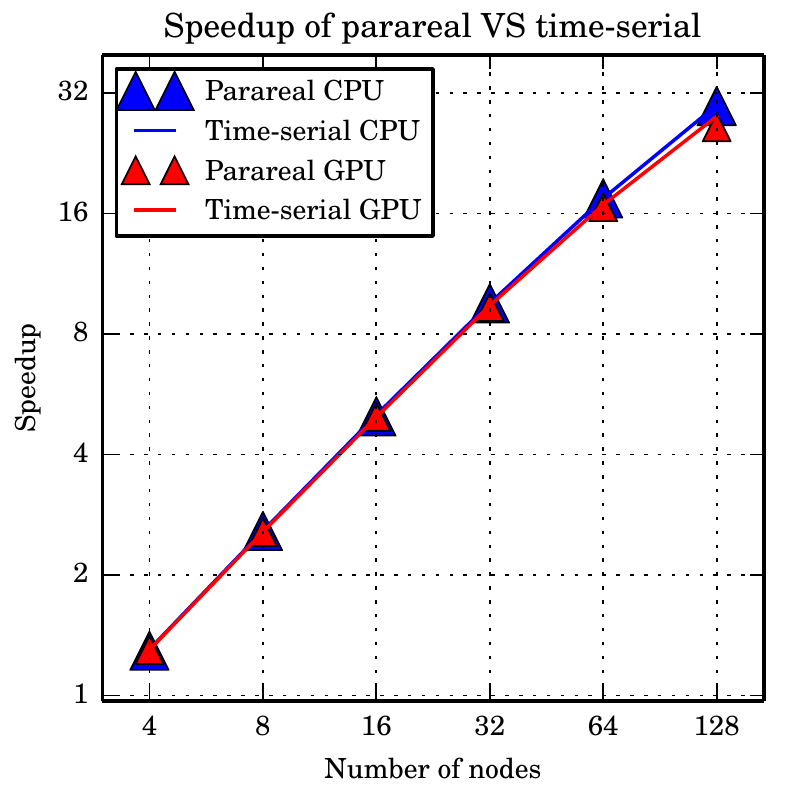}
		\caption{Speedup}\label{fig:speedup}
	\end{subfigure}	
	\caption{Runtime in seconds (left) and speedup (right) of Parareal depending on the number of nodes using the nodes' GPU (red) or CPU (blue) for stencil evaluations. Speedup is measured against $\mathcal{F}_{\delta t}$ run serially on a single node, using the CPU or the GPU, respectively. The solid lines in (b) show the maximum theoretical speedup $S_{bound}(N_p)$.}
\end{figure}
Figure~\ref{fig:runtimes} shows the actual runtimes for the CPU and GPU versions of Parareal, depending on the number of nodes.
The horizontal blue and red line indicate the runtimes of the serial reference runs.
The GPU version is constantly about a factor of  $4.5$ faster than the CPU version, but both curves are more or less parallel to each other, suggesting that achieved speedups should be similar for both backends.
This is confirmed by Figure~\ref{fig:speedup}, which shows the speedup provided by Parareal depending on the number of nodes, i.e. the number of time slices.
Red triangles denote GPU runs, blue triangles CPU runs, and the lines show the theoretical maximum speedup according to~\eqref{eq:speedup}.
Speedup is measured against a time-serial run of the fine method using a single node and STELLA's CPU or GPU backend, respectively.
Performance of Parareal in terms of speedup is nearly oblivious against the choice of the STELLA backend: Speedups are almost identical for the CPU and GPU version.

Table~\ref{tab:efficiency} summarizes speedups $S_{\rm bound}$ and $S_{\rm measured}$ as well as efficiencies $E_{\rm bound}$ and $E_{\rm measured}$ defined in~\eqref{eq:efficiency} for the CPU and GPU version of Parareal-STELLA depending on the number of nodes.
Expected speedup is computed from~\eqref{eq:speedup}, with the coarse-to-fine ratio $\tau_c / \tau_f$ measured from serial runs and therefore slightly different for both versions.
Up to $128$ nodes, the CPU version provides speedups and efficiencies very close to the theoretical bounds.
The slightly super-optimal value for $64$ nodes comes from the inaccuracy in the formula for $S_{\rm bound}$ by computing the ratio $\tau_{\rm c}/\tau_{\rm f}$ experimentally.
The GPU version also provides speedups and efficiencies very close to the theoretical bound, with a little drop off for large node numbers.
This is probably due to the fact that actual runtimes here are much smaller so that small overheads have a larger effect.
\begin{table}[t]
\centering
\begin{tabular}{|c|c|c|c|c|} \hline
\multicolumn{5}{|c|}{CPU version} \\ \hline
\# Nodes & $S_{\rm bound}$ & $S_{\rm measured}$ & $E_{\rm bound }$ & $E_{\rm measured}$ \\ \hline
4               & 1.3                          & 1.3                                 & 32.6                         & 32.3 \\ 
8               & 2.6                          &  2.6                                & 32.2                         & 32.2 \\
16             & 5.0                          &  5.0                                & 31.4                         & 31.1 \\
32             &  9.5                         &  9.5                                & 29.8                         & 29.8 \\
64             & 17.4                       &  17.6                              & 27.3                         & 27.5 \\
128           & 29.8                       &  29.7                              &  23.3                        & 23.2 \\ \hline
\end{tabular}
\begin{tabular}{|c|c|c|c|c|} \hline
\multicolumn{5}{|c|}{GPU version} \\ \hline
\# Nodes & $S_{\rm bound}$ & $S_{\rm measured}$ & $E_{\rm bound}$ & $E_{\rm measured}$ \\ \hline
4               & 1.3                          &  1.3                                & 32.5                        & 32.4 \\ 
8               & 2.6                          &  2.6                                & 32.0                        & 32.0 \\
16             & 5.0                          &  5.0                                & 31.0                        & 31.0 \\
32             & 9.4                          &  9.3                                & 29.4                        & 29.2 \\
64             & 16.8                       &  16.6                               &26.3                         & 25.9 \\
128           & 28.0                       &  26.3                               & 21.9                        & 20.5 \\ \hline
\end{tabular}
\caption{Speedup and parallel efficiency in percent of the CPU and GPU version of Parareal-STELLA depending on the number of nodes used for the time-parallelization. The theoretical bounds for speedup and efficiency according to~\eqref{eq:speedup} are denoted as $S_{\rm bound}$ and $E_{\rm bound}$.
The measured values are $S_{\rm measured}$ and $E_{\rm measured}$. The coarse-to-fine runtime ratios $\tau_{\rm c} / \tau_{\rm f}$ in~\eqref{eq:speedup} are determined experimentally by running coarse and fine method on a single node.}\label{tab:efficiency}
\end{table}

\subsubsection{Energy consumption}\label{subsec:energy}
The energy consumption of a supercomputer constitutes a large part of its effective cost. Moreover, there is an increasing concern about the impact of scientific computing on the environment and about the sustainability of the growth of the performance of supercomputers.
For these reasons, in the last decade, energy consumption has become a major issue in high-performance computing. 
A prominent sign of the increasing importance is that, besides the Top500 ranking of the most powerful supercomputers~\cite{top500}, a new ranking called Green500 publishes a list of the most energy-efficient supercomputers~\cite{green500}. Piz Daint, the machine of CSCS where we carried our experiments, entered this ranking at position four, being the most energy-efficient petaflop machine. On this machine, a comprehensive set of power management facilities has been installed by Cray~\cite{energy2} and these tools have been tested in a study to assess the energy efficiency of multiple applications~\cite{energy}. 
Assessing the performance of algorithms not only in terms of time-to-solution but also energy-to-solution is important, but for time-parallel methods there seem to be no such studies yet.

\begin{figure}[t]
	\centering
	\begin{subfigure}[t]{0.495\textwidth}
		\centering
		\includegraphics[width=0.99\textwidth]{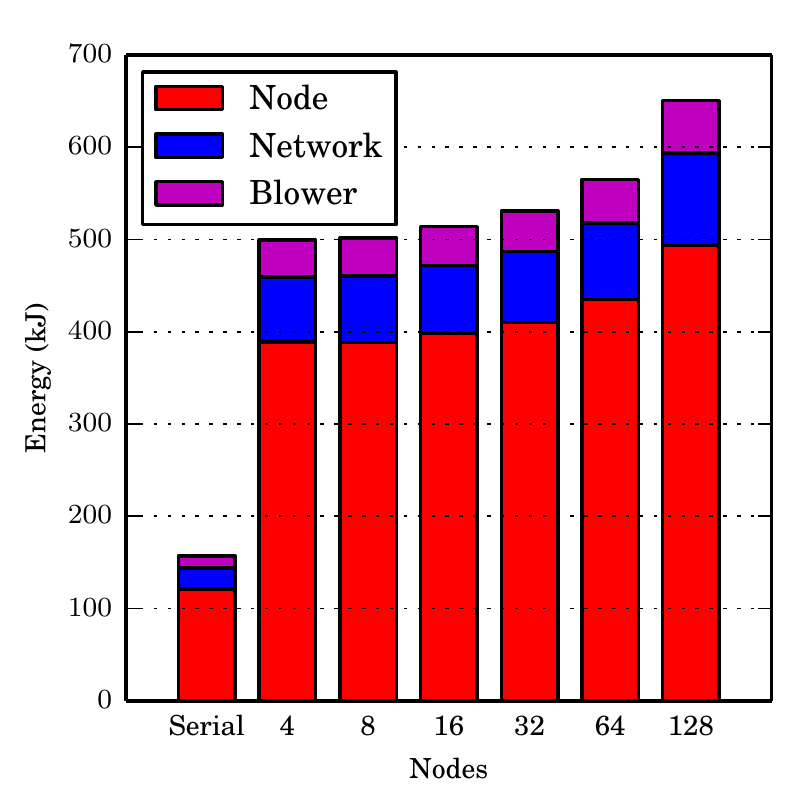}
		\caption{Energy-to-solution in Kilojoule for CPU backend.}
	\end{subfigure}
	\begin{subfigure}[t]{0.495\textwidth}
		\centering
		\includegraphics[width=0.99\textwidth]{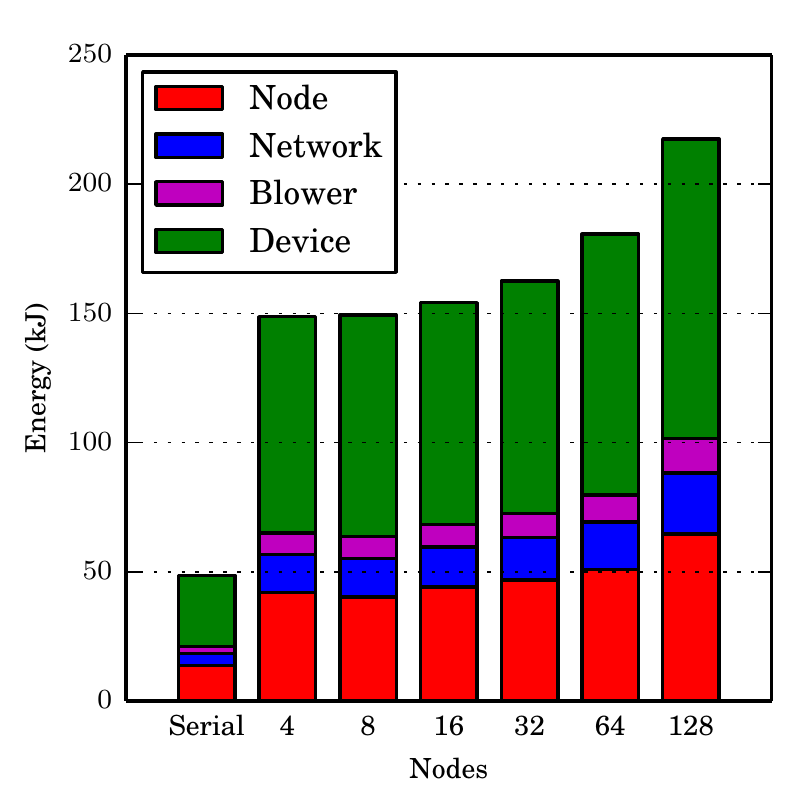}
		\caption{Energy-to-solution in Kilojoule for GPU backend.}
	\end{subfigure}
    \caption{Total energy consumed by Parareal-STELLA with CPU backend (left) and GPU backend (right) depending on the number $N_{\rm p}$ of nodes or concurrently computed time-slices. The red portion is the energy consumption of the CPU, the main memory and all other components of the compute node. The blue portion corresponds to the energy consumption of the interconnect network. The purple portion represents the consumption of the cooling system of the machine. The green portion is the consumption of the GPU, which is shown only in the GPU version.}
    \label{fig:energy}
\end{figure}

Figure~\ref{fig:energy} shows the total energy consumed by the supercomputer Piz Daint to perform the time-serial integration on a single node and the Parareal algorithm on multiple nodes.
Energy consumed by the node, i.e. mainly main memory and CPU, is marked in red.
Blue indicates energy consumed by the network while purple is energy consumed by the cooling system.
Finally, for the GPU backend, the energy consumed by the GPU itself is marked in green, and labeled as \textit{Device} energy.
The GPU backend has a factor of three to four lighter energy consumption throughout (note that differently scaled y axes).
Going from a single node (time serial) to four or more nodes (Parareal) results in a significant increase in consumed energy (about a factor of three) for both the CPU and the GPU version.
After that, the energy consumption from using more nodes for Parareal increases at a much slower pace: The reason is that e.g. from Parareal across four to Parareal across eight nodes, the runtime is reduced by about a factor of two (see e.g. Table~\ref{tab:efficiency}, the speedup on eight nodes is double the speedup on four nodes).
Therefore, while running Parareal on eight nodes doubles power consumption, because it also about halves runtime the energy-to-solution stays about the same. 
Thus, the difference in energy-to-solution between Parareal runs using different numbers of nodes is governed by Parareal's intrinsic parallel efficiency while the difference between Parareal and the serial propagator is naturally governed by Parareal's parallel efficiency in comparison to $\mathcal{F}_{\delta t}$.
For larger numbers of concurrently computed time slices, Parareal's intrinsic efficiency start to go down, too, and energy-to-solution increases again when compared with Parareal on four nodes.

In Table~\ref{tab:power} we summarize the power consumption per node of different components of the machine when performing the simulations. 
Power consumption depends heavily on the kind of computation performed, on its computational intensity, on the usage of memory and on other factors. 
Nevertheless, the runs with the time-serial integration scheme and those with Parareal showed very similar per-node power consumption, with variations on the order of a few percent.
This supports the relatively coarse assumptions made for the derivation of $\gamma_{\rm bound}$ above and suggests that Parareal's overhead in terms of energy should, as expected, be closely linked to its parallel efficiency with little impact from additional overhead e.g. from communication.

For both backends, the major portion of power (three quarters or more) is required by the computing devices, i.e the node for the OpenMP and the node and the device for the CUDA backend. The power required by the network is fixed at a rate of 25 W/node, while the cooling system constantly consumes about $14$ W/node.
It is interesting to note that the GPU backend actually has a significantly higher power consumption than the CPU version, so that its lower energy-to-solution is only due to its noticeably smaller runtimes.

\begin{figure}[t]
	\centering
	\begin{subfigure}[t]{0.495\textwidth}
		\centering
		\includegraphics[width=0.99\textwidth]{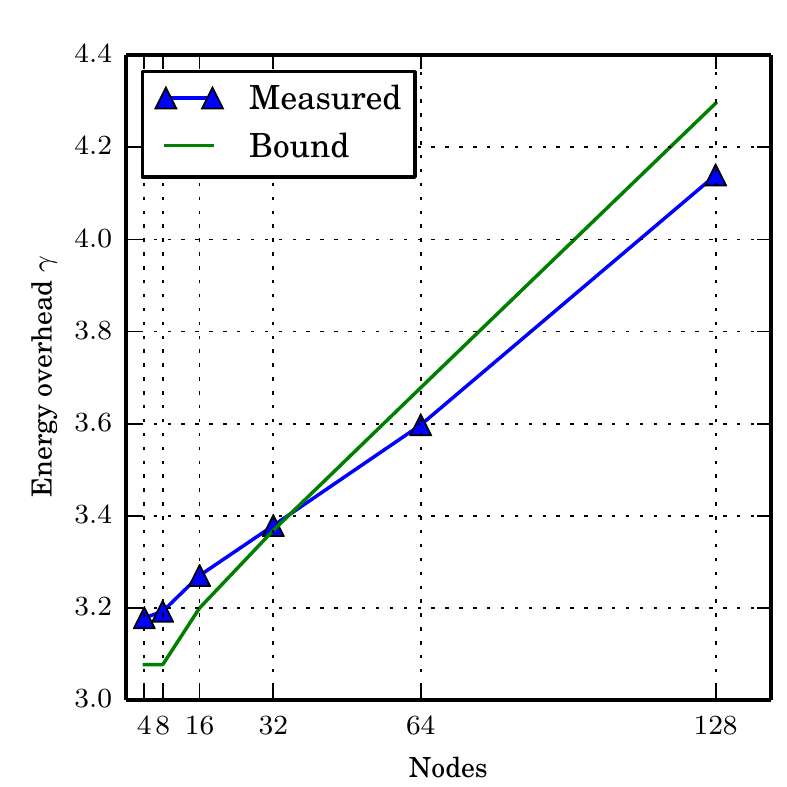}
		\caption{Energy overhead $\gamma$ for CPU backend.}
	\end{subfigure}
	\begin{subfigure}[t]{0.495\textwidth}
		\centering
		\includegraphics[width=0.99\textwidth]{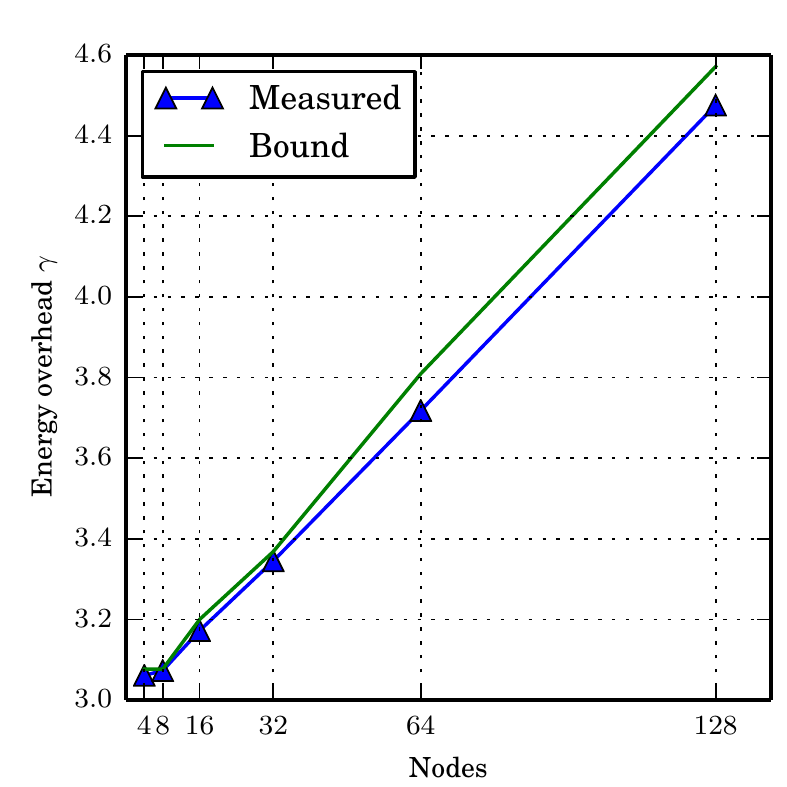}
		\caption{Energy overhead $\gamma$ for GPU backend.}
	\end{subfigure}
	\caption{Energy overhead of Parareal as defined in \eqref{eq:overhead}, i.e.\ the energy-to-solution required by Parareal divided by energy-to-solution for the time-serial fine method. The expected overhead according to~\eqref{eq:gamma_expected}, represented by the green line without markers, indicates the overhead that is unavoidable due to Parareal's intrinsically less than optimal parallel scaling. This bound contains experimentally measured terms and is thus not infinitely accurate, which is the reason for the apparent super-optimal behavior. }\label{fig:energy_overhead}
\end{figure}

Figure~\ref{fig:energy_overhead} plots the measured energy overhead of Parareal, that is
\begin{equation}
	\label{eq:overhead}
	\gamma_{\rm measured} = \frac{ Q_{\rm p} }{ Q_{\rm s} },
\end{equation}
cf.~\ref{subsubsec:speedup}.
In addition, the value $\gamma_{\rm bound}$ defined in~\eqref{eq:gamma_expected} is shown, that is the overhead that is to be expected from the sub-optimal scaling of Parareal. 
For both the CPU and the GPU backends, the actual energy overhead is for some reason slightly better than the theoretical minimum, but gives a reasonable estimate.
The reasons most likely are the inaccuracies in both the energy measurement as well as the experimental computation of the term $\tau_f / \tau_c$ in the speedup estimate and thus the theoretically expected parallel efficiency that enters into $\gamma_{\rm bound}$.
A more comprehensive analysis would have to sample over a large ensemble of runs and invoke statistical measures here, in order to provide a more detailed estimate.
This is, however, beyond the scope of the current paper.
Nevertheless, Figure~\ref{fig:energy_overhead} strongly suggests that the significant overhead of Parareal in terms of energy depicted in Figure~\ref{fig:energy} can to a large degree be attributed to its rather harsh limit on parallel efficiency: Therefore, improving the parallel efficiency of Parareal in particular but also related parallel-in-time methods in general will simultaneously significantly improve efficiency in terms of energy consumption.
\begin{table}[t]
    \centering
    \begin{tabular}{|l|c|c|c|c|}
	\hline
	\multirow{2}{*}{} & \multicolumn{2}{c|}{CPU}                                           & \multicolumn{2}{c|}{GPU}                                           \\ \cline{2-5} 
			  & \multicolumn{1}{l|}{Consumption} & \multicolumn{1}{l|}{Percentage} & \multicolumn{1}{l|}{Consumption} & \multicolumn{1}{l|}{Percentage} \\ \hline
	Node              & 133 W                            & 77 \%                           & 70 W                             & 29 \%                           \\ \hline
	Network           & 25 W                             & 15 \%                           & 25 W                             & 10 \%                           \\ \hline
	Blower            & 14 W                             & 8 \%                            & 14 W                             & 6 \%                            \\ \hline
	Device            & ---                              & ---                             & 135 W                            & 55 \%                           \\ \hline
	\textbf{Total}    & 172 W                            & 100 \%                          & 245 W                            & 100 \%                          \\ \hline
    \end{tabular}
    \caption{Power consumption per node of the considered simulations with different backends. For each component the power consumption and the its percentage of the total is shown. These values are the result of empirical measurements performed on Piz Daint. The Node and Device (GPU) power consumption fluctuates depending on the intensity of computations, while the Network and Blower power consumption is constant. The difference in power consumption between the time-serial integration and Parareal with different numbers of nodes is negligible and thus only one set of values is shown.}
    \label{tab:power}
\end{table}

\section{Conclusions and outlook}
The paper presents a stencil-based implementation of the time-parallel Parareal method in the C++ domain specific embedded language STELLA.
STELLA is the acronym for STEncil Loop Language and is designed for easy and efficient implementation of stencil-based computations on different platforms.
It provides different backends, e.g.\ using OpenMP or CUDA, for the parallelized evaluation of stencils on structured grids.
In the here presented combination, STELLA is used for the spatial intra-node parallelization, using either the node's GPU or multi-core CPU, depending on the backend.
Parareal is then added to parallelize in time across nodes, using MPI for communication of volume data within the Parareal iteration.
This corresponds to a paradigm using distributed memory parallelization in time and shared memory parallelization in space.

Performance of the stencil-based Parareal implementation is assessed in terms of runtimes, speedup and energy consumption.
A three-dimensional advection-diffusion equation with a time-dependent diffusion coefficient is used as a benchmark problem.
First, convergence of Parareal is analyzed and a setup is defined where the time-parallel solution provides accuracy comparable to the time-serial reference.
In line with previous results, it is found that a time-dependent diffusion coefficient has only marginal influence on the convergence of Parareal.
Then, the performance of both the OpenMP and CUDA backend are investigated and compared.
For both versions, measured speedups closely match theoretically predicted values, illustrating the efficiency of the approach.
Speedup from Parareal is found to be roughly identical for both backends, with the GPU backend being consistently about a factor of 4.5 faster in terms of runtimes.
In terms of energy-to-solution, Parareal is found to cause significant overhead, with the GPU version requiring significantly less energy than the CPU version for both time-serial and time-parallel runs.
To distinguish between energy overhead caused by the intrinsically sub-optimal parallel efficiency of Parareal and additional overhead e.g.\ from communication, an estimate for the energy overhead is derived from the theoretical bound on speedup and compared to the measured overhead.
It is found that the significant rise in energy-to-solution from Parareal can be attributed to a very large degree to its harsh bound on parallel efficiency.
This result is of interest for other ''parallel across the time-steps" approaches as well (e.g.\ PITA~\cite{FarhatEtAl2003}, PFASST~\cite{EmmettMinion2012} or time multigrid~\cite{FriedhoffEtAl2013} or~\cite{Neumueller2014}) because it suggests that improving parallel efficiency of such methods will also be key to improve their efficiency in terms of energy.
Derivation of a more detailed model for the energy consumption of Parareal or other time-parallel methods would a very interesting next step.

Moreover, in order to have an analytic solution available, the benchmark considered here is a relatively simple linear advection-diffusion problem.
Having been designed to be used in the dynamical core of the numerical weather prediction and climate model COSMO, STELLA can be employed for complex nonlinear equations, too.
Studying the performance of Parareal-STELLA for such a problem is planned for future work.

This paper considered spatial parallelism across one single node/GPU combined with time-parallelism across nodes.
For larger problems, where a time-slice does not fit on a single node, the Parareal-STELLA benchmarks can be extended to use distributed memory in time and a hybrid parallelization in space, similar to what has been explored for a particle discretization in~\cite{SpeckEtAl2012}.
In the other direction, for problems too small to fully utilize a single GPU or many-core CPU with only spatial parallelism, a shared-memory space-time parallel approach could extend the degree of concurrency and be used to improve utilization of a single node.
So far, this approach has not been explored, but the framework studied here would provide a good starting point for such a study.
Other studies exist that use shared memory in time on a single node~\cite{RuprechtKrause2012} or in combination with a distributed memory parallelization in space~\cite{HaynesOng2014,RuprechtKrause2014_DDM}.
Moreover, pure MPI based approaches have also been studied e.g.\ in~\cite{Trindade2006} or~\cite{RuprechtEtAl2013_SC}.
A  meticulous comparison of different implementation strategies for space-time parallelization in terms of important metrics like speedup, energy consumption, memory requirements, etc. would be a very interesting future research direction.

\section*{Acknowledgments}
The authors thankfully acknowledge the support of the Swiss PASC (``Platform for Advanced Scientific Computing)'' initiative.  They would also like to thankfully acknowledge the many helpful discussions with Oliver Fuhrer from MeteoSwiss, Carlos Osuna from C2SM and Will Sawyer, Benjamin Cumming and Gilles Fourestey from the Swiss National Supercomputing Centre CSCS.

\bibliography{Pint,Pint_Self,HPC,other_refs}







\end{document}